\documentclass[preprint,12pt]{elsarticle}

\usepackage{graphicx}
\usepackage{epstopdf}
\usepackage{amsmath,amssymb}
\usepackage{booktabs}
\usepackage{mathrsfs}
\usepackage{multirow}
\usepackage{bm}

\usepackage{dcolumn}% Align table columns on decimal point
\usepackage{longtable}
\newcommand{\be}{\begin{eqnarray}}
\newcommand{\ee}{\end{eqnarray}}

\usepackage[normalem]{ulem}  % \sout{old text} for strikeout
\usepackage[dvips]{color} % For blue in-text comments and additions
\renewcommand\sout{\bgroup \color{red} \ULdepth=-.5ex \ULset}

\biboptions{sort&compress}

\journal{}

\begin{document}

\begin{frontmatter}



\title{Exotic dibaryons with a heavy antiquark}  %\tnoteref{aaa}
\tnotetext[a]{Report No.: KEK-TH-1676, J-PARC-TH-29}


\author[a]{Yasuhiro Yamaguchi}
\ead{yamaguti@rcnp.osaka-u.ac.jp}
\author[b]{Shigehiro Yasui}
\author[a,c]{Atsushi Hosaka}

\address[a]{Research Center for Nuclear Physics (RCNP), 
Osaka University, Ibaraki, Osaka, 567-0047, Japan}
\address[b]{KEK Theory Center, Institute of Particle and Nuclear
Studies, High Energy Accelerator Research Organization, 1-1, Oho,
Ibaraki, 305-0801, Japan}
\address[c]{J-PARC Branch, KEK Theory Center, Institute of Particle and
 Nuclear Studies, KEK, Tokai, Ibaraki, 319-1106, Japan}

\begin{abstract}

We discuss the possible existence of exotic dibaryons with a heavy antiquark,
being realized as three-body systems, $\bar{D}^{(\ast)}NN$ and
 $B^{(\ast)}NN$. These are genuinely exotic states with no
 quark-antiquark annihilation.
We consider the heavy quark spin and chiral symmetries, 
and introduce the one pion exchange potential between a $\bar{D}^{(\ast)}$ ($B^{(\ast)}$) meson and a
nucleon $N$.
As for the $NN$ interaction, we employ the Argonne $v^\prime_8$ potential.
By solving the coupled-channel equations for $PNN$ and $P^\ast NN$
  ($P^{(\ast)}=\bar{D}^{(\ast)}$ and $B^{(\ast)}$),
we find bound states for $(I,J^P)=(1/2,0^-)$ as well as resonant states for
 $(I,J^P)=(1/2,1^-)$ both in $\bar{D}^{(\ast)}NN$ and $B^{(\ast)}NN$ systems.
We also discuss the heavy quark limit, 
and find that the spin degeneracy is realized in the bound states with $(I,J^P)=(1/2,0^-)$ and $(1/2,1^-)$.
\end{abstract}


\end{frontmatter}


%\section{}
%\label{}

%======================================================
\section{Introduction}
%======================================================
Many exotic hadrons observed in experiments are considered to be
loosely bound states of two hadrons, called hadronic molecules or
hadronic composites~\cite{Brambilla:2010cs}.
The study of such configurations is important 
to establish interactions among hadrons as inputs of various hadronic phenomena,
 such as formation of bound/resonant states and decay processes in
 hadronic molecules.  
Moreover, the hadron interactions provide us with useful information to study
 fundamental problems of QCD such as color confinement, chiral
symmetry breaking and so on.  

The hadron-nucleon interaction is the basic quantity, not only for hadronic molecules, but also for exotic nuclei.
In fact, the hyperon-nucleon interaction determines 
the properties of a variety of
hypernuclei~\cite{Hashimoto:2006aw,Hiyama:2010zzb,Hiyama:2010zz}.
Furthermore it is suggested by theoretical
studies~\cite{Akaishi:2002bg,Yamazaki:2003hs,Dote:2008hw,Gal:2013vx}
that the attraction between a $\bar{K}$ meson and a nucleon would lead to the
formation of $\bar{K}$-mesic nuclei, which are also researched in
experiments~\cite{Agnello:2005qj,Yamazaki:2010mu}.
In the heavy quark sector, there have been, not only analogous
discussions, but also new approaches which are not accessible in light
flavor sectors.
There the heavy quark symmetry~\cite{Isgur:1989vq,Isgur:1991wq,Flynn:1992fm} becomes important.  
Under this symmetry, an interesting observation was
made, namely there is a sufficiently strong attraction due to the tensor
force of the one pion exchange at long distances between a $\bar{D}^{(\ast)}$
$(B^{(\ast)})$ meson and a nucleon $N$ leading to the $\bar{D}^{(\ast)}N$ ($B^{(\ast)}N$) molecules
around the thresholds~\cite{Yasui:2009bz,Yamaguchi:2011xb,Yamaguchi:2011qw,Cohen:2005bx,Gamermann:2010zz}.
Here, we note that $\bar{D}^{(\ast)}$ stands for $\bar{D}$ or $\bar{D}^\ast$.
A unique 
feature of such molecules is the exotic flavor structure of the minimum quark content $\bar{Q}qqqq$, where $\bar{Q}$ is a heavy
antiquark and $q$'s are light quarks.
Hence, the $\bar{D}^{(\ast)}N$ ($B^{(\ast)}N$) molecules are genuinely
exotic baryons, having no lower hadronic channels coupled by a strong
decay.

In literatures, however,
there are other theoretical studies suggesting that repulsions at short
distances and 
weaker pion coupling may change the interaction less
attractive or even repulsive, which makes no bound state~\cite{Haidenbauer:2007jq,Carames:2012bd,Fontoura:2010sxa,Fontoura:2012mz}. 
Our strategy here is to determine the long range part of the
interaction, namely the one pion exchange potential with the strong
tensor force, with respecting the heavy quark symmetry.

The possible attraction between a $\bar{D}^{(\ast)}$ $(B^{(\ast)})$ meson and a nucleon motivates
us
to explore the few-body problems in exotic nuclei with heavy quarks,
because it is naturally expected that the binding energy becomes larger as the baryon number increases.
In the light flavor sector, it has been studied that the hadron-nucleon
interaction gives us rich phenomena in few-body systems such as the impurity
effects, e.g. shrinking of the wave functions due to glue-like
effects in hypernuclei~\cite{Hashimoto:2006aw} and possible high
density states in $\bar{K}$-mesic nuclei~\cite{Yamazaki:2003hs}, 
 which have never been realized in normal nuclei.
In the heavy flavor sector, however, few-body systems of 
$\bar{D}^{(\ast)}$ ($B^{(\ast)}$) nuclei with a few baryon numbers have not been
investigated so far in the literature,
 though there have been several works for $\bar{D}^{(\ast)}$ ($B^{(\ast)}$)
mesons in nuclear
matter with infinite volume \cite{Tsushima:1998ru,Mishra:2003se,Lutz:2005vx,Hilger:2008jg,Tolos:2007vh,Mishra:2008cd,Kumar:2010gb,JimenezTejero:2011fc,Kumar:2011ff,GarciaRecio:2011xt,Yasui:2012rw,Yasui:2013} (see Ref.~\cite{Yasui:2012rw} for a summary of recent results) 
and in atomic nuclei such as $^{12}$C, $^{40}$Ca and $^{208}$Pb with
larger baryon numbers \cite{Tsushima:1998ru,GarciaRecio:2011xt,Yasui:2012rw}.
The few-body systems would be more likely to be produced in experiments
in hadron colliders rather than the nuclei with middle and large baryon numbers.

In this paper, we study the mass spectrum of $\bar{D}^{(\ast)}NN$ and
$B^{(\ast)}NN$ bound and/or resonant states 
of dibaryons (baryon number two) with a heavy antiquark
for the first time. 
We emphasize that $\bar{D}^{(\ast)}NN$ and $B^{(\ast)}NN$ are unique
as few-body systems, because the counterparts of $KNN$ in the
strangeness sector do not exist due to the repulsive interaction between a $K$ meson and a
nucleon.
Moreover, 
 we can reveal the role of the heavy quark spin symmetry, as a
 fundamental symmetry of QCD, 
 in multi-hadron systems with a heavy quark, which does not appear
 in light flavor systems.

This paper is organized as follows.
In the next section, we briefly summarize the interaction 
of $\bar{D}^{(\ast)}N$ ($B^{(\ast)}N$) and $NN$, where we emphasize
the role of the one pion exchange potential
which has been investigated in
Refs.~\cite{Yasui:2009bz,Yamaguchi:2011xb,Yamaguchi:2011qw}.
In Sec.~\ref{sec:wave_function}, we show the method to solve the
$\bar{D}^{(\ast)}NN$ ($B^{(\ast)}NN$) systems with appropriate three-body wave functions.
In Sec.~\ref{result}, the numerical results
are shown,
where we study bound and resonant states both for $\bar{D}^{(\ast)}NN$ and $B^{(\ast)}NN$ %states
with quantum numbers $J^P=0^-$ and $1^-$, and $I=1/2$.
We also discuss the spin degeneracy of the three-body systems in the heavy
quark mass limit.
In the last section, we summarize the present work.

%======================================================
\section{Interactions}\label{int.}
%======================================================

Let us start with the discussions of the basic interaction for the $P^{(\ast)}N$
($P^{(\ast)}=\bar{D}^{(\ast)}$ and $B^{(\ast)}$).  
As we have emphasized in
Refs.~\cite{Yasui:2009bz,Yamaguchi:2011xb,Yamaguchi:2011qw} for the
system with a heavy meson 
with a heavy antiquark, the one pion exchange potential (OPEP) is 
the basic ingredient to provide a strong attraction.  
The existence of the OPEP is a robust consequence of chiral symmetry in  
the presence of a light quark in the $P^{(\ast)}$ meson, while its importance 
in the $P^{(\ast)}N$ interaction is supported by the heavy quark symmetry of
QCD~\cite{Isgur:1989vq,Isgur:1991wq,Flynn:1992fm}.  
The relevance of the heavy quark symmetry is in the spin degeneracy 
of $P$ and $P^\ast$ mesons.  
It is related to the mass degeneracy between $P$ and
$P^{\ast}$
as experimental data shows the small mass differences, 
$m_{\bar{D}^\ast}-m_{\bar{D}}\sim 140$ MeV and $m_{B^\ast}-m_{B}\sim 45$ MeV.
Thanks to these small mass splittings, 
$P$ and $P^\ast$ mesons mix dynamically 
through the couplings of  
$PN-P^\ast N$ and 
$P^\ast N-P^\ast N$ 
caused by the OPEP. 
We note that the OPEP does not exist in $PN-PN$ only, because the
$PP\pi$ vertex cannot exist from the parity conservation.
The interaction for $PN-PN$ is effectively supplied from the mixing
processes like $PN \rightarrow P^{\ast}N \rightarrow PN$. 
Thus, the OPEP together with the heavy quark spin symmetry provides a
unique dynamics in the systems of a heavy meson. 

The OPEP between a $P^{(\ast)}$ meson and
a nucleon $N$ can be described from the heavy meson effective
theory~\cite{Yasui:2009bz,Yamaguchi:2011xb,Yamaguchi:2011qw}.
The OPEPs for $PN-P^\ast N$ and $P^\ast N-P^\ast N$ are given by
\begin{align}
 &V_{PN\!-\!P^\ast N}(r)=- \frac{g_\pi g_{\pi NN}}{\sqrt{2}m_N f_\pi}\frac{1}{3} 
 \left[\vec{\varepsilon}\,^\dagger\cdot\vec{\sigma}C(r)+S_\varepsilon
 T(r)\right]\vec{\tau}_P \!\cdot\! \vec{\tau}_N, \label{PNP*N} \\
 &V_{P^\ast N\!-\!P^\ast N}(r)= 
 \frac{g_\pi g_{\pi NN}}{\sqrt{2}m_N f_\pi}\frac{1}{3}
 \left[\vec{S}\cdot\vec{\sigma}C(r)+S_S
 T(r)\right]\vec{\tau}_P \!\cdot\! \vec{\tau}_N, \label{P*NP*N}
\end{align}
as a sum of the central and tensor forces, $C(r)$ and $T(r)$,
where $m_N=940$ MeV and $f_\pi=132$ MeV are the mass of the nucleon
and the pion decay constant, respectively.
In Eqs. \eqref{PNP*N} and \eqref{P*NP*N}, 
$\vec{\varepsilon}$
($\vec{\varepsilon}^{\,\dag}$) is the polarization vector of the
incoming (outgoing) $P^\ast$,
$\vec{S}$ is the spin-one operator of $P^\ast$, and $S_\varepsilon$ ($S_S$)
is the tensor operator $S_{\cal O}(\hat{r})=3(\vec{\cal
O}\cdot\hat{r})(\vec{\sigma}\cdot\hat{r})-\vec{\cal
O}\cdot\vec{\sigma}$ with $\hat{r}=\vec{r}/r$ and $r=|\vec{r}\,|$ for
$\vec{{\cal O}}=\vec{\varepsilon}$ ($\vec{S}\,$), where $\vec{r}$ is the
relative position vector between $P^{(\ast)}$ and $N$.
$\vec{\sigma}$ are Pauli matrices acting on nucleon spin.
$\vec{\tau}_P$ ($\vec{\tau}_N$) are isospin operators for $P^{(\ast)}$ ($N$).
The coupling constant for $P^{(\ast)}P^{\ast} \pi$ vertex is given by
$g_\pi=0.59$ which is determined by the experimental value of the decay
width of $D^\ast \to D\pi$~\cite{Yasui:2009bz,Yamaguchi:2011xb,Yamaguchi:2011qw}.
Here we use $g_\pi$ with the same value for Eqs.~\eqref{PNP*N} and \eqref{P*NP*N}
from the heavy quark spin symmetry.
The coupling constant $g^2_{\pi NN}/4\pi=13.6$ for $NN\pi$ vertex
is given in Ref.~\cite{Machleidt:2000ge}.
The functions $C(r)$ and $T(r)$ are
\begin{align}
 &C(r)=\int\frac{d^3 \vec{q}}{(2\pi)^3}
 \frac{m_{\pi}^2}{\vec{q}^{\,\,2}+m_{\pi}^2}e^{i\vec{q}\cdot\vec{r}}F(\vec{q}\,)\,
 , \label{Cpote}\\
 &S_{\cal O}(\hat{r})T(r)=\int\frac{d^3 \vec{q}}{(2\pi)^3}
 \frac{-\vec{q}^{\,\,2}}{\vec{q}^{\,\,2}+m_{\pi}^2}S_{\cal O}(\hat{q})e^{i\vec{q}\cdot\vec{r}}F(\vec{q}\,)
 , \label{Tpote}
\end{align}
with $\hat{q}=\vec{q}/|\vec{q}\,|$, where the dipole-type form factor
$F(\vec{q}\,)=(\Lambda_{P}^2-m_{\pi}^2)(\Lambda_{N}^2-m_{\pi}^2)/(\Lambda_{P}^2+|\vec{q}\,|^2)(\Lambda_{N}^2+|\vec{q}\,|^2)$
with cutoff parameters $\Lambda_P$ and $\Lambda_N$ is introduced for
spatially extended hadrons. 
From a quark model estimation, 
we use $\Lambda_D=1.35\Lambda_N$ for $\bar{D}^{(\ast)}$ meson, $\Lambda_B=1.29\Lambda_N$ for
$B^{(\ast)}$ meson, and $\Lambda_{P_{Q}}=1.12\Lambda_N$ for a $P_{Q}^{(\ast)}$ meson defined as a meson $(\bar{Q}q)_{\mathrm{spin}\,0(1)}$
having an infinitely heavy quark mass, 
as discussed in Refs.~\cite{Yasui:2009bz,Yamaguchi:2011xb,Yamaguchi:2011qw}.
%, with $\Lambda_N=830$ MeV. %,
The cutoff $\Lambda_N=830$ MeV is determined to reproduce the binding energy of the deuteron, 
2.22 MeV, when only the OPEP is used as the $NN$ potential.
As we discussed in Ref.~\cite{Yamaguchi:2011xb} and further verified,
the resulting OPEP provides a reasonable one-parameter approximation to
the low energy properties of the $NN$ system in the $^3S_1-^3D_1$
channel, including the deuteron, as summarized in
Table~\ref{table:NNlowpro}.

%==========================================================
\begin{table}[h]
 \caption{Low energy properties of the $NN$ system, the binding energy
 $E_B$, relative distance of proton and neutron\protect \footnotemark[1] $\langle r^2 \rangle^{1/2}$
 and quadrupole
 moment $Q_d$ of the deuteron, and $NN$ scattering length $a$ and
 effective range $r_e$. The predictions obtained by the OPEP~\cite{Yamaguchi:2011xb} and
 AV$8^\prime$, and 
 experimental values (Exp.) summarized in Ref.~\cite{Wiringa:1994wb} are compared.}
 \label{table:NNlowpro}
 \begin{center}
  \begin{tabular}{lccccc}
   \hline
   &$E_B$ [MeV] & $\langle r^2 \rangle^{1/2}$ [fm] & $Q_d$ [fm] & $a$ [fm] &
   $r_e$ [fm]\\ \hline
   OPEP &2.22 &3.7 &0.24 &5.27 &1.50 \\
   AV$8^\prime$ &2.23 &3.9 &0.27 &5.39 &1.75 \\
   Exp. &2.22 &3.9 &0.29 &5.42 &1.76 \\
   \hline
  \end{tabular}
 \end{center}
\end{table}
 \footnotetext[1]{In Ref.~\cite{Yamaguchi:2011xb}, the quantity $\langle r^2 \rangle^{1/2}$
 should have been called a relative distance as we do in this article, 
 rather than the radius of the deuteron.  Besides this misleadingness, actual calculations were done properly.   
 }
% what we defined the root
% mean square radius $\langle r^2 \rangle^{1/2}$ is an expectation value of a relative distance $r$
% between particles, which is twice larger than the general definition, $\langle r^2 \rangle^{1/2}/2$.
% In this article, we call $\langle r^2 \rangle^{1/2}$ a relative
% distance 
%% to avoid confusion for readers.
% } 
%==========================================================

We note that, although the OPEP has been known to play an essential role
in the nucleon-nucleon interaction,
attention has not been paid in the meson-nucleon interaction.
In fact, the mixing effect is much suppressed in the light
meson sector like 
($\pi,\rho$) and ($K,K^\ast$), where the mass splittings are much
larger than in the heavy meson sector, $m_{\rho}-m_{\pi}\sim 600$ MeV
and $m_{K^\ast}-m_{K}\sim 400$ MeV, and therefore the OPEP plays only a
minor role.
We expect that the nuclear systems containing $P^{(\ast)}$ mesons, in
accordance with the
heavy quark spin symmetry, revive the importance of the role of the
OPEP in the hadron-nuclear systems.

In the present work, we employ the OPEP for the interaction of $P^{(\ast)}N$, 
while we could have a short-range interaction as provided by the $\rho$ and
$\omega$ meson exchanges, 
or even by the quark exchange
model~\cite{Haidenbauer:2007jq,Gamermann:2010zz,Carames:2012bd,Fontoura:2010sxa,Fontoura:2012mz,Hofmann:2005sw,Hofmann:2006qx}.  
Although these interactions are less determined than the OPEP, 
it turns out that they are almost irrelevant 
for $P^{(\ast)}N$ and $P^{(\ast)}NN$ systems.  
In fact, we have numerically verified in the previous study
that the two-body matrix elements of the OPEP are much larger than those 
of the short-range interaction of $\rho$ and $\omega$ exchanges; 
the latter are typically less than ten percent~\cite{Yamaguchi:2011xb} of the former, 
an analogous situation in the deuteron.  
This argument justifies the use of the OPEP for the $P^{(\ast)}N$. 

For the nucleon-nucleon interaction, though the OPEP is
reasonably good for the description of low energy properties, we employ
a more realistic potential of Argonne $v^\prime_8$ (AV8$^\prime$)~\cite{Pudliner:1997ck}.
It is written as
\begin{align}
 &v^\prime_8(r)= \sum_{p=1}^{8} v_p(r){\cal O}^{p} \, ,
\end{align}
formed by a sum of 8 operators ${\cal O}^{p=1,\cdots, 8}$; the central operators 
$[1,(\vec{\sigma}_1\cdot\vec{\sigma}_2)]\otimes[1,(\vec{\tau}_1\cdot \vec{\tau}_2)]$,
the tensor operators
$S_{12}\otimes[1,(\vec{\tau}_1\cdot \vec{\tau}_2)]$, and 
the $LS$ operators 
$(\vec{L}\cdot\vec{S})\otimes[1,(\vec{\tau}_1\cdot \vec{\tau}_2)]$.
The function $v_p(r)$ is given in Ref.~\cite{Pudliner:1997ck}.
The AV8$^\prime$ potential is simpler than the more elaborated one of
the Argonne $v_{18}$ (AV18) potential~\cite{Wiringa:1994wb},
which is the reason that we employ the former in the present study.
The AV$8^\prime$ is realistic because it reproduces $NN$ phase shifts
and deuteron properties.
When applied to three or more nucleon systems,
however, the AV$8^\prime$ provides slightly more attraction than the AV18 potential.
In the $\bar{D}^{(\ast)}NN$ systems, however, there are only two nucleons with the
binding energy similar to that of the deuteron, and therefore, the
AV$8^\prime$ gives essentially the same results as the AV18.

%======================================================
\section{Three-body wave functions}\label{sec:wave_function}
%======================================================

The total Hamiltonian is given by
\begin{align}
 &H=T+V_{P^{(\ast)}N}+V_{NN},
 \label{Hamiltonian}
\end{align}
where $T$ is the kinetic term, and $V_{P^{(\ast)}N}$ ($V_{NN}$) is
the interaction potential between a $P^{(\ast)}$ meson and a nucleon
(between two nucleons) as introduced above.
We note that the heavy quark spin symmetry is respected in the
$V_{P^{(\ast)}N}$ potential.
The small violation of the heavy quark spin symmetry is taken into account in the mass splitting between $P$ and $P^{\ast}$.
We investigate $\bar{D}^{(\ast)}NN$ and $B^{(\ast)}NN$ with $J^P=0^-$ and $1^-$ and $I=1/2$ (total angular momentum
$J$, parity $P$ and total isospin $I$). 
We also consider $P^{(\ast)}_{Q}NN$ in the heavy quark limit.

In order to express the three-body wave functions, we employ the Gaussian
expansion method~\cite{Hiyama:2003cu} which is one of the powerful
methods to solve few-body calculations and has been utilized to
investigate bound and scattering states of such systems in hadron and
nuclear physics.
In the study, we do not solve the Faddeev equation which gives accurate
solutions for the three-body systems.
However, it has been discussed that the Gaussian method works well and
that their results are equivalent to those by the 
Faddeev method, when the
good convergence of the eigenenergy is obtained~\cite{Hiyama:2003cu}.

\begin{figure}[t]
 \begin{center}
  \includegraphics[width=8cm,clip]{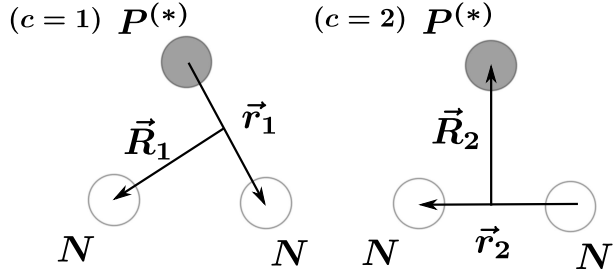}
  \caption{Jacobi coordinates of the $P^{(\ast)}NN$ systems.}
  \label{fig:jacobi}
 \end{center}
\end{figure}

The three-body wave function is described as a sum of the rearrange
channel amplitudes ($c=1,2$) as functions of the Jacobi coordinates shown in
Fig.~\ref{fig:jacobi}:
\begin{align}
 \Psi_{JM}=&\sum_{c=1}^{2}\sum_{nl_1,Nl_2,L}\sum_{s_{12}S,I_{12}} C^{(c)}_{nl_1,Nl_2,L,s_{12}S,I_{12}I}{\cal A}
 \left\{
 \left[\left[\phi^{(c)}_{nl_1m_1}(\vec{r}_c)\psi^{(c)}_{Nl_2m_2}(\vec{R}_c)\right]_{L}
 \right.\right. \notag\\
 &\times
 \left.\left.
 \left[\left[\chi_{s_1}\chi_{s_2}\right]_{s_{12}}\chi_{s_3}\right]_{S}
 \right]_{JM}
 \left[\left[\eta_{I_1}\eta_{I_2}\right]_{I_{12}}\eta_{I_3}\right]_{I}
 \right\}
 \, . \label{Eq:gaussian_expansion}
\end{align}
${\cal A}$ is the anti-symmetrization operator for exchange between
two nucleons.
$l_1$ and $l_2$ stand for the relative orbital angular momenta
associated with the coordinates $\vec{r}_c$ and $\vec{R}_c$,
respectively.
$L$ is the total orbital angular momentum of the
three-body system.
$\chi_{s_i}$ ($\eta_{I_i}$) with $i=1,2,3$ is the spin (isospin)
function of
the particle with the spin $s_i$ (isospin $I_i$).
$s_{12}$ ($I_{12}$) is the spin (isospin) of two particles combined by the
relative coordinate $\vec{r}_c$, and $S$ is the total spin of the three-body system.
The functions $\phi_{nl_1m_1}(\vec{r}\,)$ and $\psi_{Nl_2m_2}(\vec{R}\,)$ are
expressed in terms of the Gaussian functions~\cite{Hiyama:2003cu} as
%by the Gaussian expansion method~\cite{Hiyama:2003cu} as
\begin{align}
 \phi_{nl_1m_1}(\vec{r}\,)&=\sqrt{\frac{2}{\Gamma(l_1+3/2)b^3_n}}\left(\frac{r}{b_n}\right)^{l_1}\exp{\left(-\frac{r^2}{2b^2_n}\right)}Y_{l_1m_1}(\hat{r})
 \, , \\
 \psi_{N{l_2}m_2}(\vec{R}\,)&=\sqrt{\frac{2}{\Gamma({l_2}+3/2)B^3_N}}\left(\frac{R}{B_N}\right)^{l_2}\exp{\left(-\frac{R^2}{2B^2_N}\right)}Y_{l_2m_2}(\hat{R})
 \, .
\end{align}
The Gaussian ranges $b_n$ and $B_N$ are given by the form of geometric
series as
\begin{align}
 b_n&=b_1 a^{n-1}\, , \quad B_N=B_1 A^{N-1} \, .
\end{align}

%=======================================================
% Figure{Jacobi coordinate}
%=======================================================

For the sum of Eq.~\eqref{Eq:gaussian_expansion}, we include all
possible coupled channels to obtain
solutions with sufficiently good accuracy.  
For instance, we include orbital angular momentum %space 
of $l_1,l_2\leq 2$.
Furthermore,
we consider two independent isospin states to form the total isospin 
$I= 1/2$.
For instance, we include the $NN$ subsystems of $I = 0$ and 1 which are combined 
with the $\bar{D}^{(\ast)}$ meson of $I = 1/2$ for the total $I = 1/2$.  

By diagonalizing the total Hamiltonian using the three-body bases introduced above,
we obtain eigenenergies and coefficient $C^{(c)}_{nl_1,Nl_2,L,s_{12}S,I_{12}I}$. 
We also calculate the poles for resonances as complex eigenvalues by
using the complex scaling
method~\cite{Aguilar:1971-269,Balslev:1971-280,Simon:1972-1,PTP.116.1}.

%---------------------------------------------
\section{Numerical Results}
\label{result}
\begin{figure}[h]
 \begin{center}
  \includegraphics[width=8.6cm,clip]{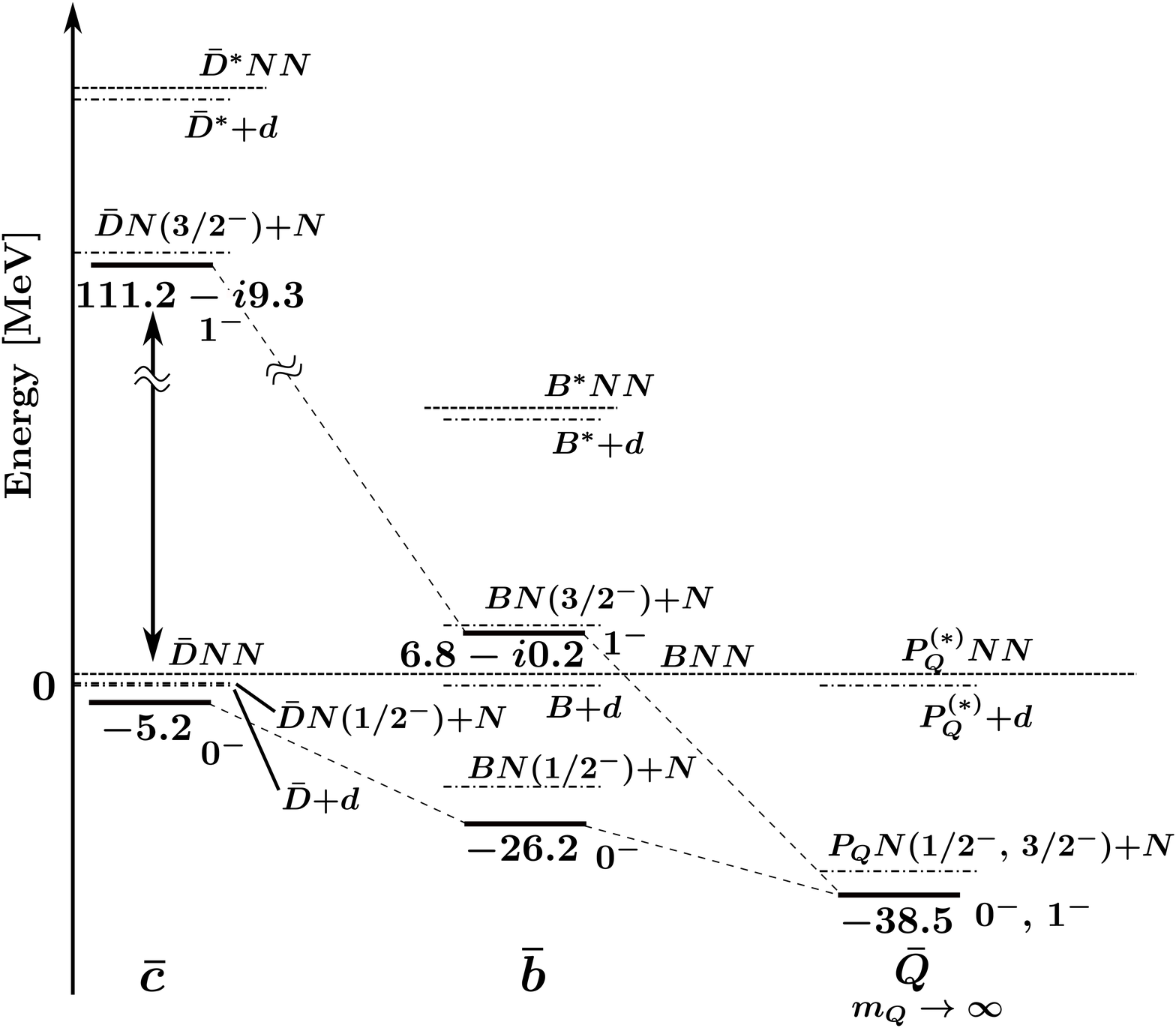}
  \caption{Energy levels of $\bar{D}^{(\ast)}NN$, $B^{(\ast)}NN$ and $P^{(\ast)}_{Q}NN$ 
  with $I=1/2$ and $J^P=0^-$ and $1^-$ (solid lines).
  The complex energies for resonances are
  given as $E_{re}-i\Gamma/2$, where $E_{re}$ is a resonance energy and
  $\Gamma/2$ is a half decay width. 
  Thresholds (subthresholds) are denoted by dashed (dashed-dotted) lines.
  }
  \label{energy-levelDbarNN}
 \end{center}
\end{figure}

%---------------------------------------------
Let us present the results of $\bar{D}^{(\ast)}NN$ and $B^{(\ast)}NN$ for $J^P=0^-$. 
We obtain bound states both of $\bar{D}^{(\ast)}NN$ and $B^{(\ast)}NN$
with energy levels shown in Fig.~\ref{energy-levelDbarNN}.
The bound state of $\bar{D}^{(\ast)}NN$, whose binding energy is $-5.2$ MeV, locates
below the threshold of $\bar{D}N(1/2^-)+N$.
%=========================================================
% Figure{Energy-level}
%=========================================================
%=========================================================
Here $\bar{D}N(1/2^-)$ is the bound state of $\bar{D}^{(\ast)}$ and $N$ with binding
energy $-1.6$ MeV for $J^P=1/2^-$ and $I=0$ as discussed in
Refs.~\cite{Yasui:2009bz,Yamaguchi:2011xb}. 
Therefore, the three-body state of $\bar{D}^{(\ast)}NN$ is more bound than
the two-body state of $\bar{D}^{(\ast)}N$, as naturally expected.
We also find the $B^{(\ast)}NN$ state with the binding energy $-26.2$ MeV.
The $B^{(\ast)}NN$ state is more bound than the $\bar{D}^{(\ast)}NN$
state, because the mixing effect between $PNN$ and $P^\ast NN$ is
enhanced, when $P$ and $P^\ast$ mesons become more degenerate.

Let us investigate how the bound states are formed.
For this purpose, we analyze various
components of interaction matrix elements. 
In Table~\ref{energyDNN}, we summarize the expectation values of the potentials, $V_{PN\!-\!P^\ast N}$,
$V_{P^\ast N\!-\!P^\ast N}$ and $V_{NN}$, sandwiched by the obtained wave functions.
In $\bar{D}^{(\ast)}NN$, we find that the tensor force of $V_{\bar{D}N\!-\!\bar{D}^\ast N}$ mixing $\bar{D}NN$ and $\bar{D}^{\ast}NN$
is the dominant contribution.
In contrast, 
$V_{\bar{D}^\ast N\!-\!\bar{D}^\ast N}$ is very small.
Thus, the tensor force of $V_{\bar{D}N\!-\!\bar{D}^\ast N}$ is
a driving force giving bound states in $\bar{D}^{(\ast)}NN$. 
We note that the strong tensor force also provides a dominant attraction
in the two-body $\bar{D}^{(\ast)}N$
systems~\cite{Yasui:2009bz,Yamaguchi:2011xb}.
The same result is also applied to $B^{(\ast)}NN$.

For $V_{NN}$, 
one may expect that the tensor force causing the $^{3}S_{1}-
^{3}\!D_{1}$ mixing could be the most dominant one, because it is a
driving force giving a deuteron $d$.
However, the tensor force in $V_{NN}$ is almost irrelevant in the present systems.
In fact, it is shown in Table~\ref{energyDNN} that the tensor force is suppressed, while the central force is rather dominant.
This is reasonable because $d$ does not exist in the main component of $\bar{D}NN$ 
due to limited combinations of quantum numbers.
It may exist in the $\bar{D}^\ast NN$ component, but the amplitude of
the $\bar{D}^\ast NN$ is small due to the excess of mass of about 140
MeV ($\sim m_{\bar{D}^\ast}-m_{\bar{D}}$).
Thus, the $NN$ interaction provides only a weak attraction.
Therefore, the tensor force mixing $\bar{D}N$ and $\bar{D}^{\ast}N$ gives the most dominant
term of attraction in the $\bar{D}^{(\ast)}NN$ state.
The same behavior is also found in the $B^{(\ast)}NN$ state as shown in Table
\ref{energyDNN}.

The quantum number $0^{-}$ of the $\bar{D}^{(\ast)}NN$ state may be
investigated in experiments through two particle correlations.
The $\bar{D}^{(\ast)}NN$ state can decay into $K+\pi\,(\mathrm{or}\,\pi\pi)+N+N$ by the weak decay of a $\bar{D}$ meson.
The absence of $d$ in the final $NN$ state will be an important signal suggesting $0^{-}$.
In contrast, if $d$ might be observed, the quantum number 
 would be $1^{-}$.
It will be also the case for $B^{(\ast)}NN$.

\begin{table}[t]
 \caption{Expectation values of central, tensor and $LS$ forces
 of the $\bar{D}^{(\ast)}N$ ($B^{(\ast)}N$) and $NN$
 potentials in the bound state of $\bar{D}^{(\ast)}NN$ ($B^{(\ast)}NN$). 
 All values are in units of MeV.}
 \label{energyDNN}
 \begin{center}  
  \begin{tabular}{c|ccc}
   \hline
   $\bar{D}^{(\ast)}NN$&$\langle V_{\bar{D}N\!-\!\bar{D}^\ast N}\rangle$
   &$\langle V_{\bar{D}^\ast N\!-\!\bar{D}^\ast N}\rangle$
   &$\langle V_{NN}\rangle$\\ \hline 
   Central&$-2.3$&$-0.1$&$-9.5$\\
   Tensor&$-47.1$&$0.7$&$-0.2$\\
   LS&---&---&$-0.03$ \\ \hline\hline
   $B^{(\ast)}NN$&$\langle V_{BN\!-\!B^\ast N}\rangle$
   &$\langle V_{B^\ast N\!-\!B^\ast N}\rangle$
   &$\langle V_{NN}\rangle$\\ \hline 
   Central&$-6.5$&$0.3$&$-11.6$\\
   Tensor&$-92.0$&$-2.7$&$-1.0$\\
   LS&---&---&$-0.1$ \\ \hline
  \end{tabular}
 \end{center}
\end{table}

In scattering states,
we find resonances for $J^P=1^-$ and $I=1/2$ as shown in
Fig.~\ref{energy-levelDbarNN}.
The resonance energy for $\bar{D}^{(\ast)}NN$ is $111.2$ MeV
measured from the threshold of $\bar{D}NN$.
The decay width is $18.6$ MeV.
We note that there are open channels of the $\bar{D}NN$ and $\bar{D}+d$ scattering states below the
resonance, and of the $\bar{D}^\ast +d$ and $\bar{D}N(3/2^-)+N$ scattering states above the resonance.
Here $\bar{D}N(3/2^-)$ is a Feshbach resonance of $\bar{D}^{(\ast)}$ and
$N$ with $J^P=3/2^-$ and $I=0$, which was found in
Ref.~\cite{Yamaguchi:2011xb}.
Those scattering states are included in the present calculation.
We obtain a resonance also for $B^{(\ast)}NN$ with much smaller resonance energy and decay width, $6.8$ MeV and $0.4$ MeV, respectively.
The mechanism of formation of the resonances is interesting.
When we ignore the $\bar{D}NN$ channel and consider only the $\bar{D}^\ast NN$
channel, 
we obtain a bound state of $\bar{D}^\ast NN$.
Hence, the $\bar{D}^\ast NN$ channel predominates.
Therefore, the obtained resonance is a Feshbach resonance for the three-body system,
as in the case of the two-body 
$\bar{D}^{}N(3/2^{-})$ system~\cite{Yamaguchi:2011xb}.
These features also hold for $B^{(\ast)}NN$.

From the above analysis, 
we see that the many features of the two-body $P^{(\ast)}N$ system
survive in the three-body $P^{(\ast)}NN$ system, because the
$P^{(\ast)}N$ interaction is the dominant force which determines the main properties of the system rather than the $NN$ interaction.

Finally, we consider $P^{(\ast)}_{Q}NN$ systems in the heavy quark limit,
where $P_{Q}$ and $P^\ast_{Q}$ are exactly degenerate in mass.
Interestingly, we find the bound states both for $J^P=0^-$ and $1^-$
with the same binging energy $-38.5$ MeV measured from $P^{(\ast)}_{Q}NN$
threshold.
Those numerical results indicate that they are degenerate.
Comparing three cases of $\bar{c}$, $\bar{b}$ and $\bar{Q}$ with $m_{Q} \rightarrow \infty$,
we see that the mass splitting between the three-body systems with $J^P=0^-$ and $1^-$
decreases as the mass of the heavy quark increases, and finally those two states
become degenerate in the heavy quark limit, as shown in Fig.~\ref{energy-levelDbarNN}.

The degeneracy in $P^{(\ast)}_{Q}NN$ systems is not accidental.
Generally, the spin degeneracy is a feature in the heavy quark limit,
 not only for a single hadron, but also for multi-hadrons, when a single heavy quark is contained.
In QCD, the spin-dependent interaction of the
 heavy quark is suppressed by the inverse of the heavy quark mass.
Thus, there should exist spin doublets (singlets) forming (non-)degenerate states with
 $j\neq 0$ ($j=0$) in the heavy quark limit.
Here $j$ is the quantum number of the contained light
 components (the brown muck \cite{Flynn:1992fm} or the spin-complex \cite{Yasui:2013vca}).
The spin degeneracy is shown, not only
from the heavy {\it quark} effective theory
\cite{Neubert:1993mb,Manohar:2000dt}, but also from the heavy {\it
hadron} effective theory \cite{Yasui:2013vca}.
In the present discussion from the heavy {\it meson} effective theory
(Eqs.~(\ref{PNP*N}) and (\ref{P*NP*N})), we have confirmed that the ground state
of three-body $P^{(\ast)}_{Q}NN$ systems exhibits the spin doublet
having $0^{-}$ and $1^{-}$. 
Both of them contain a common light component
with total angular momentum and parity $j^{\mathcal{P}}=1/2^{+}$ and
isospin $I=1/2$.

%%%%%%%%%%%%%%%%%%%%%%%%%

%======================================================
\section{Summary}
\label{Summary}
%======================================================
In this paper, we have explored the possible existence of
genuinely exotic dibaryons, $\bar{D}^{(\ast)}NN$ and $B^{(\ast)}NN$ for
charm and bottom sectors, and $P^{(\ast)}_{Q}NN$ in the heavy quark
limit.
The OPEP introduced by chiral symmetry 
and the heavy quark spin symmetry was employed between a heavy meson and a
nucleon.
As for the $NN$ interaction, we employed the Argonne $v^\prime_8$ potential.
By solving the coupled-channel equations for $PNN$ and $P^\ast NN$,
we have obtained bound states with $J^P=0^-$ and Feshbach resonances
with $J^P=1^-$ for $I=1/2$ both in $\bar{D}^{(\ast)}NN$ and $B^{(\ast)}NN$.
The tensor force of the OPEP mixing $PN$ and $P^\ast N$ plays
an important role to produce a strong attraction.
For the $P^{(\ast)}_{Q}NN$ systems in the heavy quark limit,
we have obtained bound states of both $J^P=0^-$ and $1^-$, 
which are
the spin degenerate states containing the brown muck or the spin-complex
with $j^{\mathcal{P}}=1/2^{+}$ and $I=1/2$.
This is the first study to show the degeneracy in the few-body system
with a heavy antiquark.
Hence
the bound states with $J^P=0^-$ and the resonance with
$J^P=1^-$ 
in the actual charm and bottom sectors have the common origin of the
 spin doublet from the heavy quark limit.

The $\bar{D}^{(\ast)}NN$ and $B^{(\ast)}NN$ states can be searched in experiments in hadron colliders.
The productions of the exotic hadrons will be studied in relativistic
heavy ion collisions in RHIC and
LHC~\cite{Cho:2010db}.
Furthermore, the search for the $\bar{D}^{(\ast)}NN$ would be also
carried out in J-PARC and GSI-FAIR.

%==========================================================
\section*{Acknowledgments}
%==========================================================
The authors would like to thank Dr. Y.~Kikuchi for valuable discussions and fruitful suggestions.
This work is supported in part by Grant-in-Aid for Scientific Research
on Priority Areas ``Elucidation of New Hadrons with a Variety of
Flavors(E01: 21105006)''(S.~Y. and A.~H.) from the Ministry of Education, Culture,
Sports, Science and Technology of Japan,
and Grant-in-Aid for ``JSPS Fellows(24-3518)''(Y.~Y.) from Japan Society for the Promotion of Science.




%\bibliographystyle{elsarticle-num}
%\bibliographystyle{elsarticle-harv} 

%==========================================================
%\section*{References}
%==========================================================

\end{document}